\documentclass[sigconf]{acmart}

\usepackage{graphicx}
\usepackage{enumitem}
\usepackage{tabularx}
\usepackage{ulem}
\usepackage{float}
\usepackage{soul}
\usepackage{xcolor}
\usepackage{titlesec}
\usepackage{subcaption} 
\usepackage{multirow}
\usepackage{booktabs}
\usepackage{array}
\AtBeginDocument{%
  \providecommand\BibTeX{{%
    \normalfont B\kern-0.5em{\scshape i\kern-0.25em b}\kern-0.8em\TeX}}}

\begin{document}

\title{System Test Case Design from Requirements Specifications: Insights and Challenges of Using ChatGPT
}

\author{Shreya Bhatia*}
\email{shreya20542@iiitd.ac.in}
\affiliation{%
  \institution{IIIT Delhi}
  \state{Delhi}
  \country{India}
}

\author{Tarushi Gandhi*}
\email{tarushi20579@iiitd.ac.in}
\affiliation{%
  \institution{IIIT Delhi}
  \state{Delhi}
  \country{India}
}
\author{Dhruv Kumar}
\email{dhruv.kumar@iiitd.ac.in}
\affiliation{%
  \institution{IIIT Delhi}
  \state{Delhi}
  \country{India}
}
\author{Pankaj Jalote}
\email{jalote@iiitd.ac.in}
\affiliation{%
  \institution{IIIT Delhi}
  \state{Delhi}
  \country{India}
}

\thanks{*Equal Contribution.} 

\renewcommand{\shortauthors}{Bhatia, Gandhi et al.}

\begin{abstract}
System testing is essential in any software development project to ensure that the final products meet the requirements. Creating comprehensive test cases for system testing from requirements is often challenging and time-consuming. 
This paper explores the effectiveness of using Large Language Models (LLMs) to generate test case designs from Software Requirements Specification (SRS) documents.
In this study, we collected the SRS documents of five software engineering projects containing functional and non-functional requirements, which were implemented, tested, and delivered by respective developer teams. For generating test case designs, we used ChatGPT-4o Turbo model. We employed prompt-chaining, starting with an initial context-setting prompt, followed by prompts to generate test cases for each use case. We assessed the quality of the generated test case designs through feedback from the same developer teams as mentioned above. Our experiments show that about 87 percent of the generated test cases were valid, with the remaining 13 percent either not applicable or redundant. Notably, 15 percent of the valid test cases were previously not considered by developers in their testing. We also tasked ChatGPT with identifying redundant test cases, which were subsequently validated by the respective developers to identify false positives and to uncover any redundant test cases that may have been missed by the developers themselves. This study highlights the potential of leveraging LLMs for test generation from the Requirements Specification document and also for assisting developers in quickly identifying and addressing redundancies, ultimately improving test suite quality and efficiency of the testing procedure.
\end{abstract}




\maketitle

\section{Introduction}\label{sec:intro}
A Software Requirements Specification (SRS) is a document that outlines the functional and non-functional requirements of a software system \cite{Jalote1997}. Functional requirements describe the system's behaviour under specific conditions, often in the form of use cases. A use case is a usage scenario for a piece of software, listing event steps that typically define the interactions between a user role (actor) and the system to complete a requirement.
System testing involves validating the developed software against these requirements to ensure its correctness and completeness by using a set of test cases \cite{ieee_2008_f}.

Effectiveness of any testing depends on the quality of test case designs for that testing.
Test case designs involve identifying scenarios, inputs and expected outputs to validate software functionality. For example, a test case design for a login page will include steps to enter valid credentials and successful login and also steps to enter invalid credentials and get error messages \cite{O’Regan2019}.

Generating test case designs for system testing is a critical task but can be laborious and error-prone when done manually. It requires understanding the SRS thoroughly and devising test cases to cover various scenarios adequately, including edge or exceptional scenarios. This time-consuming process may lead to missing some important test cases or having redundant test cases \cite{swain_test_2010}.

Large Language Models (LLMs), such as OpenAI's GPT \cite{ lang_gpt, NEURIPS2022_b1efde53},
have shown promise in natural language understanding and generation tasks. Leveraging LLMs for test case design generation from SRS documents can potentially automate and expedite this process, benefiting software engineers by reducing the effort required for generating test case designs and improving their quality.
Our research aims to understand the effectiveness of utilizing LLMs to generate test case designs. 


Our Research Questions are: 
\begin{itemize}
\item \textbf{RQ1: How effective are Large Language Models (LLMs) in generating comprehensive and valid test cases directly from Software Requirements Specifications (SRS) documents?}\cite{intro_prompt}. 

\item \textbf{RQ2: What proportion of test cases generated by LLMs are valid yet previously overlooked by developers, and what value do these test cases add in terms of reducing developer effort and improving test coverage?} 

\item \textbf{RQ3: To what extent can LLMs identify redundant test cases within a test suite, and how accurately do these identifications align with developer-validated redundancies?} 

\end{itemize}





\section{Methodology}\label{sec:method}
In this section, we outline the characteristics of our dataset and explain the approach used to design an effective prompt for ChatGPT. Subsequently, the designed prompt is used to generate test-case designs from requirements documents using ChatGPT. 
We also discuss the evaluation metrics used to assess the quality and effectiveness of the generated test cases.

\subsection{Dataset: Projects Used in the Study}

We assembled a dataset consisting of five Software Requirements Specification (SRS) documents from student engineering projects. These projects were developed for real clients, and were implemented and tested before delivering them to the client. Each SRS document includes functional and non-functional requirements, with functionality specified as use cases. 

The five projects varied in size, ranging from 7,000 to 12,000 lines of code, with team sizes of 3 to 4 members. Each SRS document contained between 11 and 29 use cases, catering to 3 to 4 distinct user types. The technological stack utilized in these projects commonly included HTML and ReactJS for the frontend,  Django and NodeJS at the backend, and PostgreSQL, MySQL, MongoDB as the database. Table \ref{table:SRS} provides a summary of the SRS characteristics for each project. Brief description of each project is provided below:

\begin{table}[h!]
\centering
\renewcommand{\arraystretch}{1.5}
\begin{tabular}{ >{\centering\arraybackslash}p{2cm} >{\centering\arraybackslash}p{1cm} >{\centering\arraybackslash}p{1cm} >{\centering\arraybackslash}p{1cm} >
{\centering\arraybackslash}p{0.5cm} >
{\centering\arraybackslash}p{1cm} }
\hline
\textbf{SRS} & \textbf{User/ Actor Types} & \textbf{Total Use Cases} & \textbf{Code Size (LOC)} & \textbf{Team Size} & \textbf{SRS Word Count} \\
\hline
SMP Portal & 3 & 16 & 8,000 & 4 & 2699\\
\hline
Medical Leave & 4 & 29 & 12,000 & 4 & 2653\\
\hline
Student Clubs & 3 & 13 & 7,000 & 3 & 2211\\
\hline
Ph.D Portal & 3 & 11 & 7,000 & 3 & 1154\\
\hline
Changemaking & 4 & 22 & 9,000 & 3 & 1716\\
\hline
\end{tabular}
\caption{Software Requirement Specification (SRS) Data}
\label{table:SRS}
\vspace{-3em}
\end{table}



\begin{itemize}

\item \textbf{Student Mentorship Program (SMP) Portal:} This project aimed to optimize the Student Mentorship Program at a University by by streamlining communication and management between admins, mentors, and mentees (students).

\item \textbf{Medical Leave Portal:} The objective of this project was to create a portal for UG and PG students at a University to efficiently request and manage their medical leaves. The portal streamlines the approval process by involving students, doctors, and administrators, and maintains a comprehensive database of student leave records. 

\item \textbf{Student Clubs Event Management Platform:} The goal of this project was to develop a platform that manages the scheduling and organization of events hosted by student clubs in a University. The platform allows club coordinators to propose events for approval and allows students to register for approved events. It also features dedicated pages for each club, managed by their respective coordinators.

\item \textbf{Ph.D. Management Portal:} This project focused on improving the management of academic processes related to Ph.D. programs in a University. It included functionalities for thesis assessment, comprehensive exams, annual evaluations, and stipend management. 

\item 
\textbf{Changemaking Website:} This project aimed to develop a website that supports Changemakers—young individuals with a mindset for creating positive change. The website provided tools, resources, and opportunities for these individuals to lead change initiatives. Additionally, it facilitated fundraising efforts and allowed the general public to contribute through donations.

\end{itemize}

\subsection{Prompt Design}
To generate effective test case designs, the initial step was to develop a prompting approach that would yield best test case designs for a given SRS. Drawing insights from recommended prompting patterns such as role prompting, using delimiters, specifying output formats and providing clear task descriptions \cite{chen_unleashing_2024, white_chatgpt_2023}, we experimented with two prompting approaches: 


\noindent\textbf{Approach 1 - Single prompt approach:} In this approach, we constructed one prompt comprising (1) the entire SRS (2) instructions to generate the test cases along with the specified output format for the same. Our experiments showed that the test case designs generated were not very elaborate; potentially indicating that the prompt was too long, and the task was too complex to be accomplished in one prompt. On an average, 2-3 test designs were generated per use case.


\noindent\textbf{Approach 2 - Prompt Chaining:} In this approach, we first provided the SRS in the initial prompt. This was followed by individual prompts for each use case, directing the LLM to generate test cases in a specified format. We observed a significant increase in the average number of test cases generated, with approximately 9-11 test cases per use case.

Table \ref{table:comparison} provides a comparison of test cases generated by the two approaches.



\begin{table}[h!]
\centering
\renewcommand{\arraystretch}{1.5}
\begin{tabular}{ >{\centering\arraybackslash}p{3cm} >{\centering\arraybackslash}p{2cm} >{\centering\arraybackslash}p{2cm} }
\hline
\textbf{SRS} & \textbf{Approach 1} & \textbf{Approach 2} \\
\hline
SMP Portal & 3.5 & 11.6 \\
\hline
Medical Leave Portal & 1.6 & 7.5 \\
\hline
Student Clubs Portal & 3.3 & 8.6 \\
\hline
Ph.D. Portal & 2.6 & 9.7 \\
\hline
Changemaking Website & 7.0 & 15.6 \\
\hline
\textbf{AVERAGE} & \textbf{3.6} & \textbf{10.58} \\
\hline
\end{tabular}
\caption{Comparison of Average Number of Test Cases Generated in Each Approach for all the Five SRS documents}
\label{table:comparison}
\end{table}
 
Based on these explorations, we can say that prompt chaining with the first prompt giving the SRS and subsequent prompts asking the LLM to generate test cases for each use case generates better test-case designs. 

For exploring the quality of test cases generated, therefore we selected a two-stage prompt-chaining approach for generating the test case designs from the SRS:  

\begin{enumerate}
    \item \textbf{Familiarization:} Instruct the LLM to read and comprehend the provided SRS document.\\
    \underline{Prompt 1:} \textit{You are a software engineer. You are in the first stage of the Software Development Life Cycle, where you are provided with the SRS of a Medical Leave portal. The text given below in triple quotes is the System Requirements Specification of this Portal. Go through it, and you will refer to it for answering the questions in upcoming prompts.\\SRS: '''...'''}
    \item \textbf{Test Case Generation:} Subsequently, prompt the LLM to generate test case designs for specific use cases mentioned in the SRS.\\
    \underline{Prompt 2:} \textit{Using the SRS of the Medical Leave Portal that was provided earlier, generate all possible test case designs, using Specification-Based technique, for each possible use case in a tabular format having the following 4 columns: functionality/condition to be tested, input action/input values, expected output/behavior, and additional comments.\\Use case: '''...'''}
\end{enumerate}

\subsection{Test Case Generation}
We employed ChatGPT-4o Turbo model with the prompting approach 2 mentioned above for each of the five SRS documents to generate test-case designs. First, ChatGPT familiarized itself with the SRS. Then, for each use case, it generated test cases in a standardised tabular format, detailing conditions, input actions, expected outputs, and comments, facilitating organized analysis. 
So, if there were 12 use cases in the SRS, we prompt ChatGPT a total of 13 times (12+1)---the first prompt will familiarize the LLM with the SRS, followed by 12 prompts, each focusing on generating test case designs for a specific use case scenario. 

Aditonally, to mitigate randomness in the experiment, we standardized all tests by using the default temperature setting and the same model version (GPT-4 Turbo). For each SRS, we applied our prompting approach multiple times in separate Chat windows, iteratively building a cumulative union of the test sets from each attempt. The process continued until the union of the test cases remained unchanged, indicating no new distinct test cases were being generated.

For example, running the prompt for an SRS initially generated 5 distinct test cases. A second attempt produced 7 test cases, with 4 overlapping the first set. Taking the union of these sets resulted in a total of 8 unique test cases. In the next iteration, the union of this set with test cases from the third attempt was computed, and so on. The process terminated when the size of the union set no longer increased, signifying that no additional unique test cases were being identified. Some examples of the test cases generated by this approach for an SRS are shown in Table \ref{table:test-case-design}.




\begin{table}[h!]
\centering
\renewcommand{\arraystretch}{1.5}
\begin{tabular}{ |>{\centering\arraybackslash}p{2cm}| >{\centering\arraybackslash}p{2cm}| >{\centering\arraybackslash}p{1.5cm}| >{\centering\arraybackslash}p{2cm}| }
\hline
\textbf{Functionality/ Condition to be Tested} & \textbf{Input Action/ Input Values} & \textbf{Expected Output/ Behaviour} & \textbf{Additional Comments} \\
\hline
User can log in with valid credentials & Enter valid username and password & Successful login, access granted to the portal & Verify that users with correct credentials can log in successfully \\
\hline
User cannot log in with invalid credentials & Enter invalid username or password & Unsuccessful login, access denied to the portal & Verify that users with incorrect credentials cannot log in \\
\hline
Two-Factor Authentication & Enable two-factor authentication and enter valid authentication code & Successful login, access granted to the portal & Verify that two-factor authentication works as expected \\
\hline
\end{tabular}
\caption{Example of Test Case Design}
\label{table:test-case-design}
\end{table}


\subsection{ Qualitative Evaluation of Test Cases}

To assess the quality and effectiveness of the test cases generated by ChatGPT, we conducted a qualitative evaluation with feedback from the developers who authored the SRS documents. This evaluation aimed to determine the utility, relevance, and completeness of the generated test case designs in capturing the intended functional requirements, by ensuring that the test cases are evaluated by developers with a deep understanding of the requirements and functional specifications of their respective projects.\\
\textbf{Feedback Collection:}
For every use case, the developers were asked to review the test case designs generated by ChatGPT and provide feedback for each test case using the following criteria:
\begin{enumerate}
    \item \textbf{Valid:} A test case is marked as valid if it is suitable for verifying the functionality of the use case.
    \item \textbf{Redundant:} A test case is redundant, if the condition it is testing overlaps with what other test cases have already covered, leading to the unnecessary testing of the same code segments and features.
    \item \textbf{Not implemented but Valid:} Test cases that were not implemented in the actual project but are valid and non-redundant. This feedback was crucial in understanding the potential gaps in the original test coverage and how LLMs can help overcome them. 
    \item \textbf{Not Applicable:} Test cases which have irrelevant conditions that could not be used to test the functionality. 
    \item \textbf{Missed Test Cases:} Finally, the teams were asked to mention any specific test case design conditions the LLM failed to generate but which they had used in their testing. This aspect of the evaluation was vital for identifying areas where ChatGPT's understanding might have been incomplete or inaccurate.
\end{enumerate}

\subsection{Validation of Redundant Test cases Identified by ChatGPT}

Redundant test cases are undesirable in a test suite as they lead to unnecessary duplication of effort, increased execution time, and difficulty in maintaining clarity and coverage across the test suite. Identifying and eliminating redundancies ensures a more efficient and streamlined testing process, helping developers focus on meaningful tests that provide value.

To address this, we extended our methodology to include the identification of redundant test cases among those generated by ChatGPT. Redundant test cases were analyzed in the context of the entire SRS, as identifying redundancies at the use case level does not provide meaningful insights due to overlaps across use cases. Prompting ChatGPT, we explicitly asked it to flag test cases that might overlap or repeat existing ones within the generated suite. Simultaneously, developers reviewed the GPT-generated test cases and flagged redundancies from their perspective.

We then compared the two sets of identified redundancies: those flagged by ChatGPT and those flagged by developers. For overlapping redundancies, developers validated ChatGPT’s findings, categorizing them as either valid redundancies or false positives. Additionally, ChatGPT flagged new redundant test cases that were missed by developers. This process provided insights into the effectiveness of LLMs in identifying redundancies and their potential to enhance developer workflows by quickly pinpointing inefficiencies in test suites.

\section{Results}\label{sec:results}



\textbf{Answering RQ1: How effective are Large Language Models (LLMs) in generating comprehensive and valid test cases directly from Software Requirements Specifications (SRS) documents?}

We observed that on average ChatGPT generated 10–11 test cases per use case on average, using a two-stage prompt-chaining approach. Based on the feedback gathered from developers, we found that about 72.5 percent of the generated test cases were valid and had been incorporated by the developers in their actual testing process. Additionally, 15.2 percent of the generated test cases were such that the developers had not even considered previously and turned out to be valid test conditions. Adding up these two percentages, in total, 87.7 percent of the LLM-generated test cases were classified as valid test conditions. Among the remaining test cases, 9.7 percent were deemed not applicable due to irrelevance or changes in system requirements, or even because ChatGPT generated unnecessary security checks for a secure intranet environment. It was noted that ChatGPT has a limited understanding of system behaviours and implementation details based solely on the provided SRS, which in turn was also the reason behind the LLM missing out on some test conditions. The missed tests, however, were identified to be rare, only 2-3 per SRS.
Additionally, 2.6 percent of the tests were identified as redundant. Inspection of the redundant test cases revealed that ChatGPT is unable to identify overlapping steps (features) across multiple use cases. Hence, it ends up generating redundant test designs for the same steps when encountered in a new use case. For example, in the case of Medical Leave portal, the particular use case of forwarding the leave request to UG Chair or PG Chair has the same steps for two actors (a doctor and an admin). Being a common requirement for both, this scenario is independent of the actor. But ChatGPT is not able to figure this out and hence, it generated the same test twice. ChatGPT also gives multiple test cases for input validation/error handling, which can effectively be combined into one test case. Note that these redundant test cases were identified by developers in the very first test suite generated by ChatGPT. Later we specifically prompt the LLM to identify the redundant test cases and validate the findings while answering research question 3.

These results, shown in table IV, highlight the LLM’s
capability to produce a majority of valid test cases, with a
small proportion of suggestions falling into the categories of
not implemented, not applicable, or redundant.

\textbf{RQ2: What proportion of test cases generated by LLMs are valid yet previously overlooked by developers, and what value do these test cases add in terms of reducing developer effort and improving test coverage?}

Among all valid test cases generated, 15.2 percent were classified as “Not implemented but valid” by developers. These test cases addressed scenarios that were either deprioritized or not initially considered during by developers during their original testing efforts, marking them as new and valid contributions. Specifically, these new test cases contributed to:
\begin{itemize}
\item \textbf{User experience improvements}, such as reminders, user feedback mechanisms, and restoration of version history.

\item \textbf{Accessibility features}, like screen reader compatibility and voice-to-text support.

\item \textbf{Security enhancements}, including multi-factor authentication and validation checks.
\end{itemize}
Developers attributed the omission of these test case scenarios to time constraints or the low perceived priority of these features during initial testing. These additions helped to fill gaps in test coverage, ensuring that critical but previously overlooked aspects of functionality were tested. The developers acknowledged that these test cases added value by identifying edge cases and design considerations that were essential for improving the overall quality of the software.

\textbf{RQ 3: To what extent can LLMs identify redundant test cases within a test suite, and how accurately do these identifications align with developer-validated redundancies?}

ChatGPT identified an average of 12.82 percent of the total test cases as redundant per SRS, compared to 8.3 percent of developer-flagged redundancies. Upon validation, the following patterns were observed:
\begin{itemize}
\item \textbf{47.19 percent of the redundancies identified by ChatGPT were also flagged by developers}, indicating significant overlap and alignment between the two sets.

\item \textbf{22.65 percent of the redundancies identified by ChatGPT were new}, i.e., they were not flagged by developers initially but were later validated as valid redundant test cases by the developers.

\item \textbf{30.16 percent of the redundancies identified by ChatGPT were false positives}, meaning they were flagged as redundant but were actually essential for ensuring complete test coverage.
\end{itemize}

This analysis demonstrates ChatGPT's capability to effectively assist developers in identifying redundancies while also uncovering redundancies that might have been overlooked. However, the relatively high rate of false positives indicates room for improvement in the model’s understanding of context and overlapping functionalities. For example, ChatGPT sometimes flagged test cases with similar functionality across different actors (e.g., forwarding leave requests by a doctor versus an admin) as redundant, even though they were necessary to validate distinct roles.

\begin{table}[h!]
\centering
\renewcommand{\arraystretch}{1.5}
\begin{tabular}{ >{\centering\arraybackslash}p{2cm} >{\centering\arraybackslash}p{1cm} >{\centering\arraybackslash}p{1cm} >{\centering\arraybackslash}p{1cm} >{\centering\arraybackslash}p{1cm} >{\centering\arraybackslash}p{1cm} }
\hline
\textbf{SRS} & \textbf{\% Valid and implemented } & \textbf{\% Not implemented but valid} & \textbf{\% Not Applicable} & \textbf{\% Redundant} & \textbf{No. of missed tests per SRS} \\
\hline
SMP Portal & 65.62 & 24.52 & 7.23 & 2.63 & 1 \\
\hline
Medical Leave Portal & 81.78 & 9.13 & 6.39 & 2.7 & 7 \\
\hline
Student Clubs Portal & 73.74 & 14.89 & 8.76 & 2.61 & 2 \\
\hline
Ph.D. Portal & 61.72 & 15.88 & 19.6 & 2.8 & 0 \\
\hline
Changemaking Website & 79.6 & 11.5 & 6.4 & 2.5 & 1 \\
\hline
\textbf{AVERAGE} & \textbf{72.492} & \textbf{15.184} & \textbf{9.676} & \textbf{2.648} & \textbf{2.2} \\
\hline
\end{tabular}
\caption{Comparison of Test Case Coverage for Different SRS}
\label{table:results}
\end{table}

\section{Related Work}\label{sec:rw}

\textbf{Traditional Approaches}: Generating test cases from SRS documents traditionally involves manual techniques \cite{Singh_2011}, keyword-based methods \cite{keyword-rel, keyword-rel-2}, and model-based approaches \cite{model-1, noauthor_model-based_nodate, abdeen_approach_2023}, as well as other advanced methods \cite{raamesh_reliable_2010, wang_automatic_2020, zhang_systematic_2014}. However, these approaches often suffer from limitations such as scalability and subjectivity bias.

\textbf{Recent Advances in LLMs}: The advent of large language models (LLMs), such as ChatGPT, has introduced a promising alternative for automated test case generation. LLMs have demonstrated proficiency in natural language understanding and code generation tasks, which can be leveraged to transform test scenario extraction from textual requirements.

\textbf{Prompt Engineering}: Prompt engineering is crucial for generating effective and efficient responses from LLMs. Several studies \cite{chen_unleashing_2024, white_chatgpt_2023, gu_systematic_2023, kong_better_2024, NEURIPS2022_9d560961} have cataloged prompt patterns and utilized foundational principles like role-prompting, prompt-chaining, and chain-of-thought prompting to optimize LLM efficacy. We applied these techniques in our experiments, refining prompts to identify the most effective ones for our task.

\textbf{LLMs in Software Development}: Lin et al. \cite{lin_when_2024} explored the use of LLMs for generating code as part of the software development process. Their work emphasizes the importance of pre-processing and fine-tuning LLMs on software-related data to develop a robust code generation pipeline that leverages LLMs' semantic understanding.

\textbf{Automated User Story Generation}: Rahman et al. \cite{rahman_automated_2024} investigated automated generation of user stories from textual requirements using sophisticated natural language processing techniques. Their study received positive feedback from developers using the RUST questionnaire, highlighting the potential of LLMs in this domain.

\textbf{Test Case Generation using LLMs}: Several studies have focused on using LLMs for generating system test case inputs and automating different stages of software testing activities: Wang et al.\cite{10.1109/TSE.2024.3368208}conducted a comprehensive survey on the application of LLMs in software testing. Mathur et al.\cite{10112971} proposed the use of T5 and GPT-3 for automated test case generation. Luu et al.\cite{Luu2023CanCA} reported on the use of LLMs like ChatGPT for metamorphic testing. Yu et al.\cite{DBLP:conf/qrs/YuFLWC23} explored the challenges, capabilities, and opportunities of using LLMs for test script generation and migration.

\textbf{LLMs in Unit Test Generation}: LLMs have also been employed to generate unit tests from source code. Studies such as those by Sapozhnikov et al. \cite{10.1145/3639478.3640024} and El Haji et al. \cite{10556390} demonstrated the utility of LLMs in generating readable and comprehensive unit test cases. There are also studies on how LLMs can be used to generate unit test cases which will help the programmer by generating test cases that are more readable and cover a variety of edge cases \cite{bhatia_unit_2024, unit-2, unit-3, unit-4}.

\textbf{Our Contribution}: While existing studies have explored various aspects of LLMs in software testing, our work differs by providing a focused and comprehensive evaluation of generated test cases from SRS documents. We incorporate feedback directly from developers who authored and implemented the SRS, classifying the generated test cases into five categories: (1) Valid cases, (2) Redundant conditions, (3) Not implemented but valid cases, (4) Not Applicable, and (5) Missed test cases (detailed in Section \ref{sec:method} and Section \ref{sec:results}).

\section{Conclusion and Future Work}\label{sec:conclusion}
Developing test cases from requirements for system testing is a critical task which impacts the quality of the final software. Manual generation of these test cases can be time-consuming and error-prone. In this paper, we report the results of a study on using Large Language Models (LLMs) for generating test case designs from Software Requirements Specification (SRS) documents. 

In our experiment, we utilized five completed software projects for which Software Requirements Specifications (SRSs) were available. Acquiring requirements documentation for real-world engineering projects is inherently challenging due to their highly confidential nature, in addition to interviewing and gathering feedback from the developer teams at each step of experimentation and categorization of the LLM-generated results.  While we acknowledge that the dataset is small and may not sufficiently generalize the results, this work serves as an initial step toward exploring the potential of generating test cases from requirements documents. Furthermore, it lays the groundwork for identifying areas of future improvements aimed at enhancing the reliability of these test cases.  

We utilized a two-step prompting approach to guide the LLM towards generating the test cases. In the first, the LLM is familiarized with the SRS. In the second stage, prompts ask it to generate test cases for each use case.  This prompting technique was superior to the technique of asking the LLM to generate test cases for the entire SRS together. 

Our experiments show that LLMs were able to generate good test case designs, and a vast majority of them were valid. It was also able to generate a significant number of test cases that were missed by the developers earlier.

One limitation of the generated test cases is the presence of redundant test conditions. Perhaps more targeted prompting may be able to reduce these. \cite{chen_unleashing_2024, white_chatgpt_2023, gu_systematic_2023, kong_better_2024, NEURIPS2022_9d560961}; 
Additionally, our redundancy analysis emphasizes the role of LLMs in refining test case suites by identifying both overlooked and redundant conditions. But a point to note is the existence of some False positives in ChatGPT-identified redundant test cases which if removed, could reduce the effectiveness of the test suites. AI tools are not yet advanced enough to be used for testing purposes as a sole entity but can be used to assist and reduce software developers' efforts to a great extent. We plan to extend the study to the comparison of multiple LLMs and gather a more extensive dataset of Requirements Specification Documents. Future work will focus on improving the precision of LLMs in redundancy detection, potentially through fine-tuning on software engineering datasets or integrating system-level context. By addressing false positives and enhancing contextual awareness, LLMs could become even more effective in supporting efficient and high-quality system testing.
To improve upon ChatGPT's limited understanding of system behaviours and design specifications, we could additionally use the Architecture Design document of the engineering project. Such a document would describe the components and design specifications required to support the application, and incorporation of these details could enhance the robustness of test case generation.



\bibliographystyle{ACM-Reference-Format}
\bibliography{references}


\end{document}